\documentclass[pr,notitlepage,
nobibnotes,twocolumn,superscriptaddress]{revtex4-2}
\usepackage[english]{babel}
\usepackage{mathtools}
\usepackage{graphicx}
\usepackage{mathrsfs}
\usepackage{amssymb}
\usepackage{xcolor}
\usepackage{amsmath}
\usepackage{bm}
\usepackage{physics}

\usepackage[unicode=true,colorlinks=true,citecolor=blue,urlcolor=blue]{hyperref}

\begin{document}
\title{Intraband circular photogalvanic effect in Weyl semimetals}

\author{L.~E.~Golub} 
\affiliation{Physics Department, University of Regensburg, 93040 Regensburg, Germany}	
\affiliation{Ioffe Institute, 194021 St.~Petersburg, Russia}
\author{E.~L.~Ivchenko} 	
\affiliation{Ioffe Institute, 194021 St.~Petersburg, Russia}

\begin{abstract}
We apply the semiclassical theory including 
the Berry curvature dipole, side jumps and skew scattering for
a quantitative description of the  circular photogalvanic effect (CPGE) in Weyl semimetals at intraband absorption. 
In contrast to gapped systems where 
they completely exhaust all contributions to 
the CPGE current,
all previously known semiclassical 
mechanisms
give a result different from that obtained using a complete quantum-mechanical approach. We show that this difference in the existing quasiclassical and full quantum-mechanical approaches persists at all spatial ranges of the disorder potential.
Apparently, the implementation of another microscopic mechanism into the quasiclassical description of the CPGE is required.
\end{abstract}

\maketitle

\section{Introduction}

Weyl semimetals are characterized by bulk band crossings in their electronic structures, leading to gapless excitations and topologically non-trivial features that give rise to a number of exotic physical phenomena. The low-energy quasiparticles in these materials -- Weyl fermions -- are condensed-matter analogues of Weyl's chiral relativistic fermions from high-energy physics. These fermions are associated with a divergent Berry curvature, effectively acting as magnetic monopoles in the momentum space.
A hallmark of Weyl semimetals is the presence of Fermi arc surface states, which are topologically protected and connect projections of Weyl nodes of opposite chirality. In addition, these materials exhibit unusual transport and optical responses due to their topological nature.
From an application perspective, the Weyl semimetals are promising due to their topologically protected band crossings and Berry curvature singularities, which allow access to novel regimes in semiconductor physics. In particular, the gapless spectrum makes them particularly suitable for terahertz detection and other optoelectronic applications~\cite{Nagaosa2020,Lv2021}.
A new surge of interest in Weyl semimetals has appeared in recent years due to proposals to generate Weyl points near the Fermi level in strained tellurium crystals~\cite{Ideue2019,Glazov2022} -- a system where now both transport and optoelectronic effects are intensively studied~\cite{Moldavskaya2023,Golub2023}, as well as in strained HgTe~\cite{Ketkar2024}.

Interaction of Weyl fermions with electromagnetic radiation has many unexpected properties. Absorption of circularly-polarized light in a Weyl semimetal results in a dc electric current flow with a direction reversed at switching the circular polarization --
Circular Photogalvanic Effect (CPGE)~\cite{Golub2018,Golub2020}. 
CPGE has fascinating properties in both low and highly doped Weyl semimetals, where the Fermi energy $\varepsilon_{\rm F}$ is smaller and larger than the photon energy $\hbar\omega$, respectively. 
In the former case, the generation rate of the CPGE current in each Weyl node is ``quantized'', i.e. it is determined by fundamental constants~\cite{Juan2017,Rees2020}. 
In the opposite limit $\hbar\omega \ll \varepsilon_{\rm F}$, where intraband optical transitions are responsible for light absorption, the CPGE current value is proportional to $1/\omega$ according to the dimensional analysis~\cite{Nagaosa2020}. Except for the light frequency, the intraband CPGE response is determined by fundamental constants only:
\begin{equation}
\label{j_general}
\bm j = \gamma {\mathcal C e^3 \over h^2\omega} \bm \varkappa \abs{\bm E}^2.
\end{equation}
Here $\mathcal C$ is the topological charge of the Weyl node, $e<0$ is the elementary charge, $h=2\pi\hbar$ is the Planck constant, $\bm \varkappa$ is the photon momentum, and ${\bm E}$
is the electric field amplitude of the light wave. The dimensionless parameter $\gamma$ of the order of unity is determined by microscopic properties of Weyl fermions and will be calculated below.

 There are two different but equivalent approaches to calculate the constant $\gamma$ in the frequency range 
\begin{equation}
\label{freq_range}
\hbar/\tau \ll \hbar \omega \ll \varepsilon_{\rm F},
 \end{equation} 
 where $\tau$ is the transport relaxation time of Weyl fermions. In the first approach, the electric field of radiation is considered as a classical ac force, and the Boltzmann kinetic equation is used for calculation of the CPGE current. However, in this case, the classical kinetic equation should be supplemented by quantum corrections accounting for chiral properties of the Weyl fermions. These semiclassical corrections are known to be (i)~the Berry curvature dipole, (ii)~side-jumps at scattering events, and (iii)~skew scattering~\cite{Deyo,Koenig2017,Golub2020,Du2021}.
 On the other hand, interaction with light can be treated as quantum-mechanical absorption of photons. This intraband optical absorption, assisted by scattering on defects or phonons, results in the CPGE current as well. In the frequency range~\eqref{freq_range}, the two approaches are equivalent and must give identical results. This equivalence has been established in wide-band gap semiconductors, where calculations using the two approaches actually yield the same CPGE current~\cite{Golub2020}.
 The comparison of these two methods in the case of gapless Weyl semimetals is the aim of the present work.
 
 We consider a Weyl node with the topological charge $\mathcal C =\pm 1$. The Hamiltonian is characterized by the Weyl fermion velocity $v_0$ as follows:
 \begin{equation}
\label{H}
\mathcal H =\mathcal C \hbar v_0 \bm \sigma \cdot \bm k,
\end{equation}
 where $\bm k$ is the wavevector referred to the Weyl node, and $\bm \sigma$ is a vector of the Pauli matrices. The conduction band states have the energy $\varepsilon_k=\hbar v_0 k$ 
and the velocity ${\bm v}_{\bm k} = {\cal C} v_0 {\bm k}/k$, 
 and we denote the eigenspinors of the Hamiltonian corresponding to the conduction band as $u_{\bm k}$.
In the first part of the paper  we assume a short-range intra-node scattering potential $U(\bm r) = U_0 \sum_i \delta(\bm r - \bm R_i)$, where $\bm R_i$ are coordinates of scatterers distributed with a uniform density $n_i$.
The scattering matrix element is given by
 \begin{equation}
\label{U}
U_{\bm k' \bm k}=U_0\braket{u_{\bm k'}}{u_{\bm k}}.
\end{equation}
 The electric field of the light wave is taken as $\bm E_\omega(t)=\bm E \exp(-i\omega t) + \bm E^*\exp(i\omega t)$.
 The photon angular momentum $\bm \varkappa$ is related with the complex amplitude $\bm E$ via $\bm \varkappa \abs{\bm E}^2=i[\bm E \times \bm E^*]$.
 
The paper is organized as follows. In the next two Sections we present calculations of the CPGE current in the semiclassical and quantum-mechanical approaches. In Sec.~\ref{disc} we discuss the obtained results, and Sec.~\ref{summary} summarizes the work.
 
\section{Semiclassical approach}

The high rotational symmetry of the Weyl semimetal model~\eqref{H} prevents the skew scattering, see Ref.~\cite{Koenig2017}
$$\gamma_{\rm skew}=0,$$ and only the Berry curvature dipole and side-jumps contribute to semiclassical CPGE.

In the semiclassical approach, the electric current density is expressed via the distribution function of Weyl fermions in the conduction band, $f_{\bm k}$, and their velocity, $\bm v_{\bm k}$, as follows
\begin{equation}
\label{j_semicl}
\bm j^{\rm SC} = e\sum_{\bm k} \bm v_{\bm k} f_{\bm k}.
\end{equation}
The distribution function is found from the kinetic equation
\begin{equation}
\label{kin_eq}
\partial_t f_{\bm k} + {e\over \hbar}\bm E_\omega(t) \cdot \bm \nabla_{\bm k}f_{\bm k}= \sum_{\bm k'} \qty(W_{\bm k \bm k'}f_{\bm k'} - W_{\bm k' \bm k}f_{\bm k}).
\end{equation}
Here $W_{\bm k' \bm k}$ is the scattering probability between the electron states $\bm k$ and $\bm k'$.

\subsection{Berry curvature dipole (BCD) induced CPGE}

Weyl fermions have a nonzero Berry curvature $\bm \Omega_{\bm k}=i\bm \nabla_{\bm k}\times 
\expval{\bm \nabla_{\bm k}}{u_{\bm k}}$.
For the conduction band states of the Hamiltonian~\eqref{H} it reads $\bm \Omega_{\bm k}=-\mathcal C \bm k/(2k^3)$~\cite{Dantas2021,Bednik2024}. Note that 
in Eq.~(6) of Ref.~\cite{Nagaosa2020}  the signs of the conduction- and valence-band Berry curvatures should be reversed.
This means that the velocity has an anomalous contribution linear in the electric field
\begin{equation}
\delta {\bm v}_{\bm k} \equiv \bm v_{\bm k}^{\rm a}= {e\over \hbar}[\bm \Omega_{\bm k} \times \bm E(t)].
\end{equation}
For calculation of the current quadratic in the electric field, 
according to Eq.~\eqref{j_semicl} we need to substitute ${\bm v}^{\rm a}_{\bm k}$ instead of ${\bm v}_{\bm k}$ and a correction to the distribution function
linear in ${\bm E}_{\omega}(t)$.
It is given by 
\begin{equation}
\label{f1}
f_{\bm k}^{(1)}=-i {ev_0f_0'(\varepsilon_k) \over \omega k}(\bm E\cdot \bm k){\rm e}^{-i\omega t} + c.c., 
\end{equation}
where $f_0(\varepsilon)$ is the Fermi-Dirac distribution function and the prime means differentiation over $\varepsilon_k$.
Here we neglected a contribution of the collision integral owing to the inequality $\omega\tau \gg 1$.
From time-averaging of Eq.~\eqref{j_semicl} with $\delta \bm v_{\bm k}=\bm v_{\bm k}^{\rm a}$ and $f_{\bm k}=f_{\bm k}^{(1)}$ we obtain
\begin{equation}
\bm j^{\rm BCD} = \bm \varkappa \abs{\bm E}^2 {\mathcal C e^3 v_0 \over 3\hbar \omega}\sum_{\bm k} {f_0'(\varepsilon_k)\over k^2}.
\end{equation}
Calculation of the sum over ${\bm k}$ yields Eq.~\eqref{j_general} with the constant $\gamma$ given by~\cite{Dantas2021,Bednik2024}
\begin{equation}
\gamma_{\rm BCD}=-2/3.
\end{equation}

\subsection{Side-jump contribution}

Electron wavepackets shift in real space during scattering events, and the elementary shift called the side-jump is given by~\cite{Belinicher1982}
\begin{equation}
\bm r_{\bm k' \bm k} = -\qty(\bm \nabla_{\bm k'}+\bm \nabla_{\bm k}) {\rm arg}\qty(U_{\bm k' \bm k}) + \bm A_{\bm k'} - \bm A_{\bm k},
\end{equation}
where $\bm A_{\bm k} = i\braket{u_{\bm k}}{\bm \nabla_{\bm k}u_{\bm k}}$ is the Berry connection, and $U_{\bm k' \bm k}$
is the scattering matrix element~\eqref{U}.

The effect of side-jumps on electron transport is twofold: the side-jumps contribute to the velocity accumulation and change the collision integral~\cite{Deyo,Du2021}. However, although the velocity accumulation is relevant to the current driven by the linear polarization of the light, it does not contribute to CPGE~\cite{Deyo,Golub2020}. 
Therefore the side-jump induced CPGE current is given by Eq.~\eqref{j_semicl} with the band velocity $\bm v_{\bm k}=\bm \nabla_{\bm k}\varepsilon_k/\hbar$ and the quadratic in $\bm E$ correction to the distribution function $f_{\bm k}^{(2)}$ found with account for side-jumps. 

The side-jumps modify the scattering probability rate to $W_{\bm k' \bm k}=W_{\bm k' \bm k}^{(0)}+W_{\bm k' \bm k}^{\rm sj}$, where
$W^{(0)}_{\bm k' \bm k}=(2\pi / \hbar)n_i \abs{U_{\bm k' \bm k}}^2\delta(\varepsilon_k-\varepsilon_{k'})$
is the conventional Fermi's golden rule formula,
and the correction reads~\cite{Golub2020,Ortix2021}
\begin{equation}
W^{\rm sj}_{\bm k' \bm k}={2\pi \over \hbar}n_i \abs{U_{\bm k' \bm k}}^2e\bm E(t)\cdot \bm r_{\bm k' \bm k} \partial_{\varepsilon_k}\delta(\varepsilon_k-\varepsilon_{k'}).
\end{equation}
Since $W^{\rm sj}_{\bm k' \bm k}$ is linear in the electric field, and we are searching for the distribution function correction $f_{\bm k}^{(2)}$ quadratic in $\bm E$, we should take into account once the field term in the kinetic Eq.~\eqref{kin_eq}. There are two ways to make this: first account for the field term and then for side-jumps in the collision integral or, vice versa, first account for side-jumps on the equilibrium distribution, and then account for the field term~\cite{Golub2020}. 
The side-jump in the Weyl semimetal with the Hamiltonian~\eqref{H} is given by~\cite{Koenig2017}
\begin{equation}
\label{sj}
\abs{U_{\bm k' \bm k}}^2\bm r_{\bm k' \bm k} = \abs{U_0}^2 [\bm k' \times \bm k] {k'+k\over 4k'^2k^2}.
\end{equation}
It follows from this expression that the side-jump correction to the collision integral vanishes at the equilibrium distribution: the right-hand side of Eq.~\eqref{kin_eq} vanishes for $W_{\bm k' \bm k}=W^{\rm sj}_{\bm k' \bm k}$ and $f_{\bm k}=f_0(\varepsilon_k)$.
Therefore only the distribution found from the following equation gives a contribution to the side-jump induced current:
\begin{equation}
\label{eq_f2}
{f_{\bm k}^{(2)} \over \tau} = \sum_{\bm k'} W^{\rm sj}_{\bm k' \bm k} 
\qty(f_{\bm k'}^{(1)}-f_{\bm k}^{(1)}).
\end{equation}
Here $1/\tau = \sum_{\bm k'}W_{\bm k' \bm k}^{(0)}\qty[1-\bm k \cdot \bm k'/(kk')]$ is the energy-dependent transport scattering rate.

In the model of the Hamiltonian~\eqref{H} we have
\begin{equation}
\label{Uquad_tau}
\abs{U_{\bm k' \bm k}}^2 = {\abs{U_0}^2\over 2}\qty(1+{\bm k \cdot \bm k' \over kk'}),  
\quad {1\over \tau}={2\pi \over 3\hbar}n_i \abs{U_0}^2 \nu(\varepsilon_k),
\end{equation}
where $\nu(\varepsilon_k)\propto \varepsilon_k^2$ is the density of states.
Substituting $f_{\bm k'}^{(1)}$ from Eq.~\eqref{f1} into Eq.~\eqref{eq_f2}, we find 
\begin{equation}
\label{f_2}
f_{\bm k}^{(2)} = {\bm \varkappa \cdot \bm k\over k}\abs{\bm E}^2 
{e^2 \hbar v_0^2 \over \omega \varepsilon_k \nu(\varepsilon_k)} \qty[(\nu f_0')' - {1\over 2\varepsilon_k}\nu f_0'].
\end{equation}
The second term in square brackets
comes from differentiation of the side-jump~\eqref{sj} which depends on energies as $\bm r_{\bm k' \bm k} \propto (1/\varepsilon_{k'}+1/\varepsilon_k)$.

Substitution $f_{\bm k}^{(2)}$ into Eq.~\eqref{j_semicl} with the velocity $\bm v_{\bm k}=v_0\bm k/k$ gives the side-jump contribution to the current in the form of Eq.~\eqref{j_general} with
\begin{equation}
\label{gamma_sj_SC}
\gamma_{\rm sj}=-1/3.
\end{equation}
Note that the first and second terms in the square brackets in Eq.~\eqref{f_2} give contributions to $\gamma_{\rm sj}$ equal to 
$-2/3$ and $1/3$, respectively. As a result, while the first one equals to the BCD contribution, the second term reduces the total semiclassical CPGE current amplitude.
The first term in Eq.~\eqref{f_2} is presented in Eq.~(D3) of Ref.~\cite{Koenig2017} (up to a sign) but the second term is missing there.

\section{Quantum-mechanical approach}

\begin{figure}[t]
	\centering \includegraphics[width=\linewidth]{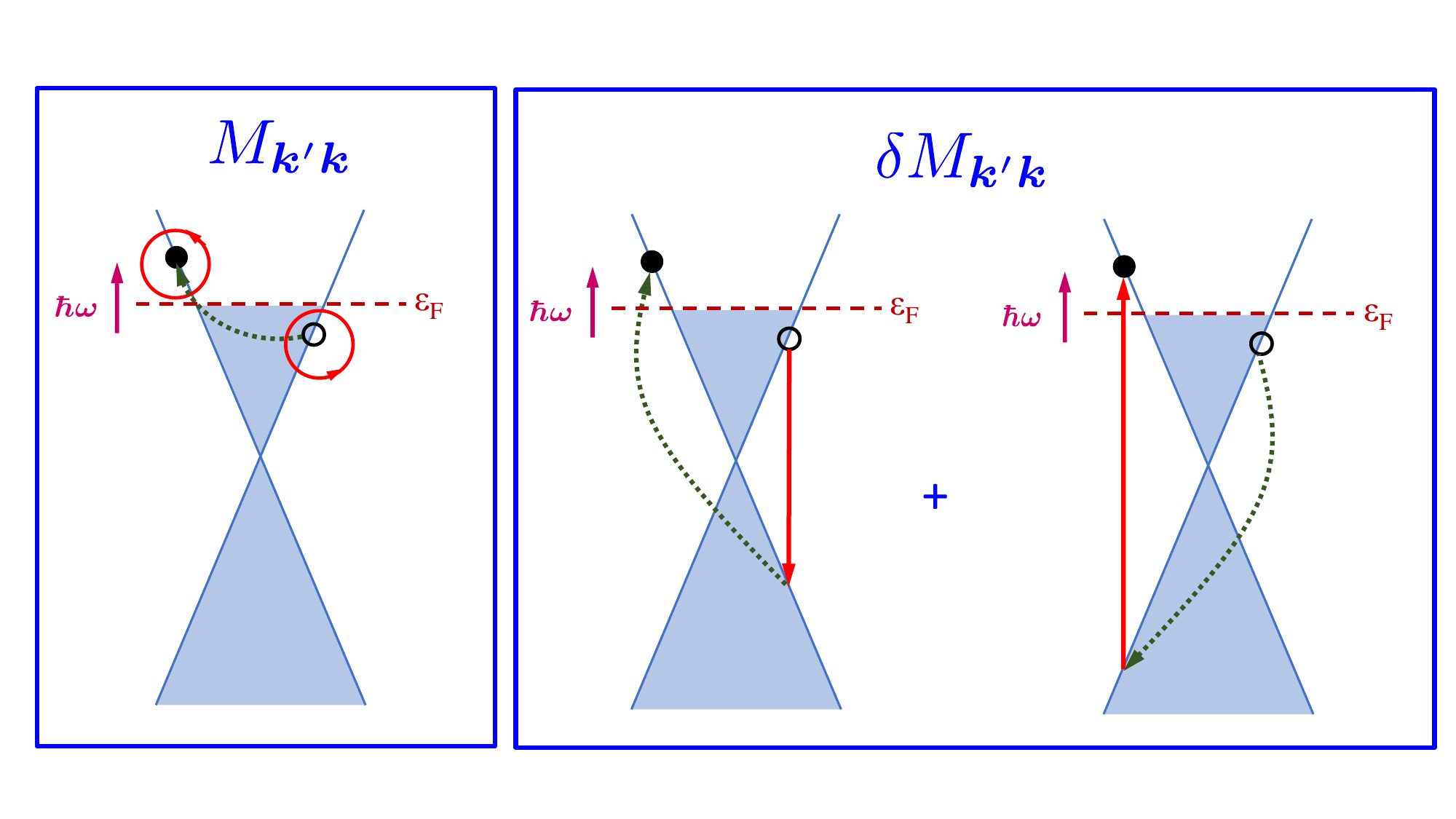} 
	\caption{ 
	Matrix elements of intraband optical transitions at $\hbar \omega \ll \varepsilon_{\rm F}$. Solid and dashed arrows denote the electron-photon interaction and momentum scattering, respectively. The matrix element $M_{\bm k' \bm k}$ shows optical transitions with intermediate states in the conduction band, and $\delta M_{\bm k' \bm k}$ describes virtual processes via the valence band states.
	}
	\label{Fig_QM}
\end{figure}

In quantum mechanics, the intraband optical absorption is described by indirect optical transitions $\bm k \to \bm k'$ with $\varepsilon_{k'}=\varepsilon_k + \hbar\omega$, assisted by momentum scattering. 
In the Weyl semimetals, 
the intermediate virtual states in these transitions are
both in the conduction band and in the valence band, Fig.~\ref{Fig_QM}.
Correspondingly, the matrix element $\mathcal M_{\bm k' \bm k}$ describing the intraband transition probability
$W_{\bm k' \bm k}^{\rm in}=(2\pi/\hbar)\abs{\mathcal M_{\bm k' \bm k}}^2\delta(\varepsilon_{k'}-\varepsilon_k - \hbar\omega)$ is given by a sum $\mathcal M_{\bm k' \bm k} = M_{\bm k' \bm k} + \delta M_{\bm k' \bm k}$.
Here the first term is a matrix element of the two-stage process with intermediate states in the conduction band:
\begin{equation}
M_{\bm k' \bm k} = {U_{\bm k' \bm k}(V_{\bm k}-V_{\bm k'}) \over \hbar \omega},
\end{equation}
where $V_{\bm k}=\expval{V}{u_{\bm k}}$ is the matrix element of the electron-photon interaction $V=ie/(\hbar\omega)\bm E \cdot \bm \nabla_{\bm k}\mathcal H$, and $U_{\bm k' \bm k}$ is the scattering matrix element~\eqref{U}.
The transitions via the valence-band states involve interband matrix elements of the electron-photon interaction and disorder scattering~\cite{Golub2018}:
\begin{equation}
\delta M_{\bm k' \bm k} = {U^{cv}_{\bm k' \bm k}V^{vc}_{\bm k}+V^{cv}_{\bm k'}U^{vc}_{\bm k' \bm k} \over 2\varepsilon_k + \hbar \omega}.
\end{equation}
Here we 
used the electron-hole symmetry relation
implying that the energy spectrum in the valence band $\varepsilon^v_k=-\varepsilon_k$.

The CPGE current density is given by~\cite{Golub2018,Golub2020}
\begin{equation}
\label{j_QM}
\bm j^{\rm QM} = e\sum_{\bm k, \bm k'} W_{\bm k' \bm k}^{\rm in} \qty[f_0(\varepsilon_k) - f_0(\varepsilon_{k'})] \qty[\bm v_{\bm k'}\tau(\varepsilon_{k'})-\bm v_{\bm k}\tau(\varepsilon_{k})],
\end{equation}
where  $\bm v_{\bm k}=\bm \nabla_{\bm k}\varepsilon_k/\hbar$ is the band velocity, 
and the energy-dependent transport relaxation time $\tau(\varepsilon)$ is given by 
Eq.~\eqref{Uquad_tau}.

The energy denominator in $\delta M_{\bm k' \bm k}$ is by a factor $\varepsilon_{\rm F}/\hbar\omega\gg 1$ larger than in $M_{\bm k' \bm k}$, therefore the dominant contribution to the light absorption is given by the processes described by $\abs{M_{\bm k' \bm k}}^2$. State of affairs with CPGE is different: in the sum
\begin{equation}
\label{M_quad}
\abs{\mathcal M_{\bm k' \bm k}}^2 = \abs{M_{\bm k' \bm k}}^2 + \abs{\delta M_{\bm k' \bm k}}^2 + 2{\rm Re}\qty({M_{\bm k' \bm k}^*\delta M_{\bm k' \bm k}})
\end{equation}
the first term makes no contribution to CPGE, while the contributions to the current~\eqref{j_QM} from the second and thirs terms are comparable.
The contribution from $\abs{\delta M_{\bm k' \bm k}}^2$ has been calculated in Ref.~\cite{Golub2018}. It is given by Eq.~\eqref{j_general} with the constant
\begin{equation}
\label{gamma_QM_1}
\gamma_{\rm QM}^{(1)}=4/3.
\end{equation}
It turns out however that, in addition to $\gamma^{(1)}_{\rm QM}$, the interference term in Eq.~\eqref{M_quad} indeed leads to the circular  photocurrent. Moreover, as shown in Appendix~\ref{App:QM2}, the ``interference'' contribution is
\begin{equation}
\gamma_{\rm QM}^{(2)}=-4/3
\end{equation}
and completely compensates the current~\eqref{gamma_QM_1}. In other words, the quantum-mechanical approach gives rise to no intraband circular photogalvanic effect.

\section{Discussion}
\label{disc}

Both semiclassical and quantum-mechanical approaches give the intraband  CPGE current in the Weyl semimetal node 
of the form of Eq.~\eqref{j_general}. 
For the parameter $\gamma$ the semiclassical treatment gives 
\begin{equation}
\label{gamma_SC_isotr}
\gamma_{\rm SC}=\gamma_{\rm BCD}+\gamma_{\rm sj}=-1,
\end{equation}
whereas the quantum-mechanical result reads
\begin{equation}
\gamma_{\rm QM}=\gamma_{\rm QM}^{(1)}+\gamma_{\rm QM}^{(2)}=0.
\end{equation}
We note that the zero value of $\gamma_{\rm SC}$ was obtained in Ref.~\cite{Koenig2017} 
as a result of inaccuracy discussed following 
Eq.~\eqref{gamma_sj_SC}, therefore it cannot be used for comparison with the quantum-mechanical value $\gamma_{\rm QM}$.

The semiclassical mechanisms of intraband CPGE considered above (BCD, side-jumps, skew scattering) are relevant for both linear and nonlinear transport. Besides, there is the interband-coherence effect
which is absent in the linear semiclassical kinetic theory~\cite{Xiao2019}. Similarly to the side-jump mechanism, it represents an effect of the electric field on scattering, but on the  matrix element rather than on the energy conservation law: the elastic scattering probability has a linear in $\bm E$ correction $\delta^{\bm E}W_{\bm k' \bm k} = (2\pi / \hbar) n_i \abs{U_0}^2\delta(\varepsilon_{k'}-\varepsilon_k) \delta^{\bm E}
\abs{\braket{u_{\bm k'}}{u_{\bm k}}}^2$
due to a $\bm E$-linear change of the overlap $\braket{u_{\bm k'}}{u_{\bm k}}$ in the scattering matrix element~\eqref{U}. 
However, as shown in Appendix~\ref{App:ib-coherence}, in the Weyl semimetals 
$\delta^{\bm E}\abs{\braket{u_{\bm k'}}{u_{\bm k}}}^2=0$, and this mechanism does not contribute to the CPGE.

We have generalized both the semiclassical and quantum-mechanical approaches to the case of angle-dependent, or anisotropic, elastic scattering. We show in Appendix~\ref{App:anis_scatt}, that the equations for CPGE are obtained from those derived for isotropic scattering by a substitution 
$n_i\abs{U_0}^2 \to \mathcal K(\abs{\bm k'-\bm k}),$ 
where $\mathcal K(q)$ is the Fourier-component of the disorder potential correlator.
The result comes down to
\begin{equation}
\label{SC_anisotr}
\gamma_{\rm SC}=-1 - {2\over 3}\varepsilon_{\rm F} {\expval{\partial_{\varepsilon_{k'}}\mathcal K(q_{\bm k'\bm k}) \sin^2\Theta}\over \expval{\mathcal K(q_\Theta)\sin^2{\Theta}}},
\end{equation}
where averaging is performed over the Fermi sphere $k=k'=k_{\rm F}$, $\Theta$ is the angle between $\bm k$ and $\bm k'$, $q_{\bm k'\bm k}=\sqrt{k'^2+k^2-2kk'\cos\Theta}$,  and $q_\Theta=\sqrt{2}k_{\rm F}\sqrt{1-\cos\Theta}$.
In the quantum-mechanical approach we have
\begin{align}
\abs{\delta M_{\bm k' \bm k}}^2 &=\mathcal K(\abs{\bm k'-\bm k})\abs{ev_0\bm E\over \omega}^2{\Psi\over 8\varepsilon_k^2},
\\
M_{\bm k' \bm k}^*\delta M_{\bm k' \bm k} &= -\mathcal K(\abs{\bm k'-\bm k})\abs{ev_0\bm E\over \omega}^2{\Phi\over 4\varepsilon_k \hbar \omega},\end{align}
where the functions $\Psi$ and $\Phi$ depend on the directions of $\bm k$ and $\bm k'$ only. Their explicit forms are given in Appendix~\ref{App:anis_scatt}.
Then we obtain from Eq.~\eqref{j_QM} the CPGE current density~\eqref{j_general} with the constant $\gamma_{\rm QM}=\gamma_{\rm QM}^{(1)}+\gamma_{\rm QM}^{(2)},$
where
\begin{equation}
\label{QM1_anisotr}
\gamma_{\rm QM}^{(1)} = {\expval{\mathcal K(q_{\bm k'\bm k})(\cos{\theta_{\bm k'}}-\cos{\theta_{\bm k}})\Psi}
\over 2\expval{\mathcal K(q_\Theta)\sin^2{\Theta}}},
\end{equation}
and
\begin{equation}
\label{QM2_anisotr}
\gamma_{\rm QM}^{(2)} = 2{\expval{\mathcal K(q_{\bm k'\bm k})(\cos{\theta_{\bm k'}}+\cos{\theta_{\bm k}})\Phi}
\over \expval{\mathcal K(q_\Theta)\sin^2{\Theta}}}.
\end{equation}
Here $\theta_{\bm k}$ and $\theta_{\bm k'}$ are the polar angles
of the vectors $\bm k$ and $\bm k'$.

\begin{figure}[h]
	\centering 
		\includegraphics[width=0.95\linewidth]{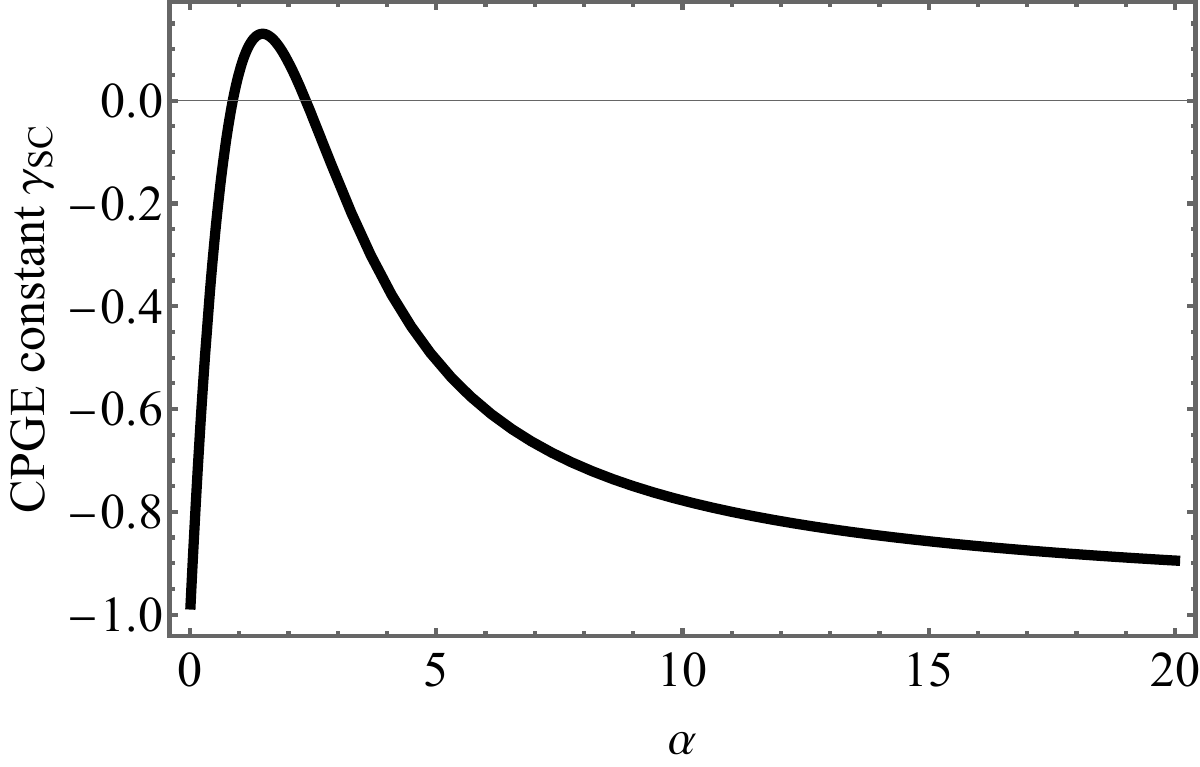}
		\includegraphics[width=0.95\linewidth]{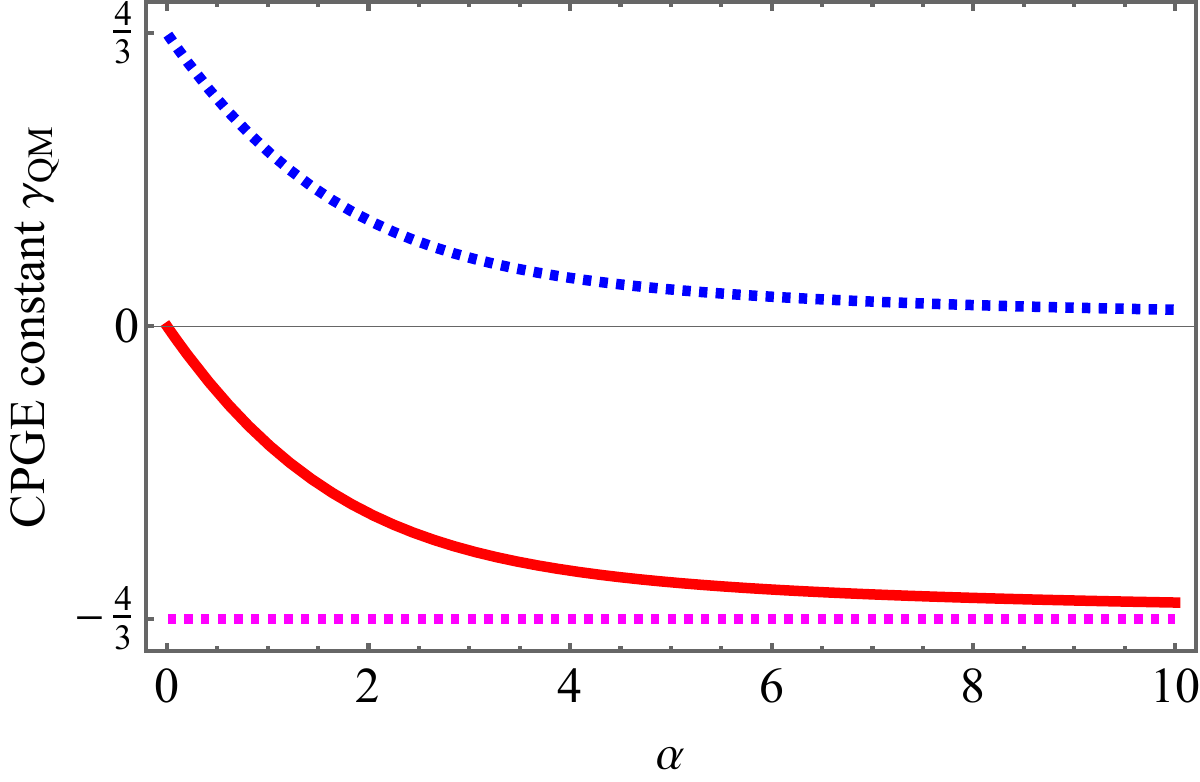}
	\caption{
The dependence of the CPGE parameter $\gamma$, Eq.~\eqref{j_general}, 
on the parameter of the Gaussian correlator $\alpha= 2(k_{\rm F}d)^2$.
Upper panel: Semiclassical result, Eq.~\eqref{SC_anisotr}.
Lower panel: the result of the quantum-mechanical calculation, $\gamma_{\rm QM}=\gamma_{\rm QM}^{(1)}+\gamma_{\rm QM}^{(2)}$ (solid curve).
Dashed curves:  the contributions $\gamma_{\rm QM}^{(1)}$, Eq.~\eqref{QM1_anisotr} (blue), and $\gamma_{\rm QM}^{(2)}$, Eq.~\eqref{QM2_anisotr} (magenta).
	}
	\label{Fig_Gauss_scatt}
\end{figure}

We consider anisotropic scattering with a Gaussian correlator $\mathcal K(q)=\mathcal K_0 \exp(-q^2d^2)$, 
and study the dependence of the CPGE current on the spatial range of the disorder potential, $d$. We obtain
the dependencies $\gamma_{\rm SC}(\alpha)$ and $\gamma_{\rm QM}(\alpha)$, where $\alpha = 2(k_{\rm F}d)^2$, see Appendix~\ref{App:anis_scatt} for details. 
They are plotted in Fig.~\ref{Fig_Gauss_scatt}.
It shows that the CPGE current in the QM approach is finite at any $\alpha > 0$ 
and vanishes only at isotropic scattering where $\alpha=0$.
Two contributions $\gamma_{\rm QM}^{(1,2)}$ display different behaviors as a function of $\alpha$: while $\gamma_{\rm QM}^{(1)}$ drops from 4/3 to 0 with increasing $\alpha$, $\gamma_{\rm QM}^{(2)}=-4/3$ is a constant. As a result, they do not cancel each other at any final $\alpha$. The total constant $\gamma$ changes from zero at $\alpha=0$ upto $-4/3$ at $\alpha \to \infty$.
In contrast, the semiclassical CPGE constant $\gamma_{\rm SC}$ exhibits a nonmonotonic dependence on $\alpha$ with a minimum of the absolute value at $\alpha\approx 1.5$ and maxima $\abs{\gamma_{\rm SC}}=1$ at $\alpha=0$ and $\alpha \to \infty$. 
Figure~\ref{Fig_Gauss_scatt} demonstrates that the two approaches give different CPGE currents at any $\alpha$.

We see that, in contrast to wide-band gap gyrotropic materials, where the equivalence of these two approaches was established in Ref.~\cite{Golub2020}, 
and despite the equivalence of the photocurrents sensitive to the linear polarization orientation (so-called Nonlinear Hall effect~\cite{Jia2024}), our quantum-mechanical and semiclassical considerations give unequal results for CPGE in the Weyl semimetals.

\section{Summary}
\label{summary}

To summarize, CPGE current is calculated in Weyl semimetals in the frequency range $\hbar/\tau \ll \hbar \omega \ll \varepsilon_{\rm F}$. The electric photocurrent is obtained  by two approaches, semiclassically and quantum-mechanically, which are expected to give identical results.  We have demonstrated that, however, these two calculation methods give different values of the CPGE currents. This indicates
an existence of extra contributions to the CPGE in gapless systems, in addition to those due to the Berry curvature dipole, side-jumps and skew scattering. 

\acknowledgments
Work of L.E.G. is supported by the Deutsche
Forschungsgemeinschaft (DFG, German Research
Foundation) via Project-ID 
521083032 (Ga501/19).

\appendix


\section{Calculation of the ``interference'' quantum-mechanical contribution}
\label{App:QM2}

We consider the Hamiltonian~\eqref{H} with the Weyl node chirality $\mathcal C=+1$ and take the envelopes of the Bloch wavefunctions in the conduction band, $u_{\bm k}$, and in the valence band, $u_{\bm k}^v$, as follows:
\begin{equation}
\label{envelopes}
u_{\bm k} = 
\begin{pmatrix}
{\rm e}^{-i\varphi/2}\cos{\theta\over 2} \\
{\rm e}^{i\varphi/2}\sin{\theta\over 2} 
\end{pmatrix},
\quad
u_{\bm k}^v= 
\begin{pmatrix}
{\rm e}^{-i\varphi/2}\sin{\theta\over 2} \\
-{\rm e}^{i\varphi/2}\cos{\theta\over 2} 
\end{pmatrix},
\end{equation}
where 
$\theta \equiv \theta_{\bm k}$ and $\varphi \equiv \varphi_{\bm k}$
are the polar and azimuthal angles of the vector $\bm k$, respectively.

The electron-photon interaction $V=(iev_0/\omega)\bm E \cdot \bm \sigma$ has the following intraband matrix elements
\begin{equation}
V_{\bm k}^{\sigma^\pm}=\expval{V^{\sigma^\pm}}{u_{\bm k}} = {iev_0\abs{\bm E}\over \omega\sqrt{2}}\sin{\theta}{\rm e}^{\pm i\varphi},
\end{equation}
where the upper and lower signs correspond to the right- and left circular polarizations: $\bm E^{\sigma^\pm}=\abs{\bm E}(1,\pm i,0)/\sqrt{2}$, and we choose the polar axis of the spherical coordinate
system directed along $\bm \varkappa$. 
The interband matrix elements read
\begin{equation}
V^{cv}_{\bm k}({\sigma^\pm})=\bra{u_{\bm k}} V^{\sigma^{\pm}}\ket{u_{\bm k}^v} = - {iev_0\abs{\bm E}\over \omega\sqrt{2}}(\cos{\theta}\pm 1){\rm e}^{\pm i\varphi},
\end{equation}
\begin{equation}
V^{vc}_{\bm k}({\sigma^\pm})=\bra{u^v_{\bm k}} V^{\sigma^{\pm}}\ket{u_{\bm k}} = - {iev_0\abs{\bm E}\over \omega\sqrt{2}}(\cos{\theta}\mp 1){\rm e}^{\pm i\varphi}.
\end{equation}

Disorder scattering matrix elements are given by
\begin{equation}
U_{\bm k' \bm k}=U_0\braket{u_{\bm k'}}{u_{\bm k}}=U_0 \qty(cc'{\rm e}^{i\phi/2} + ss'{\rm e}^{-i\phi/2})
\end{equation}
for the intraband transitions, and by
\begin{equation}
U^{cv}_{\bm k' \bm k}=U_0\braket{u_{\bm k'}}{u^v_{\bm k}}=U_0 \qty(sc'{\rm e}^{i\phi/2} - cs'{\rm e}^{-i\phi/2}),
\end{equation}
\begin{equation}
U^{vc}_{\bm k' \bm k}=U_0\braket{u^v_{\bm k'}}{u_{\bm k}}=U_0 \qty(cs'{\rm e}^{i\phi/2} - sc'{\rm e}^{-i\phi/2}), 
\end{equation}
for the interband transitions.
Here we introduced the phase $\phi=\varphi_{\bm k'}-\varphi_{\bm k}$, and the trigonometric functions $c=\cos(\theta_{\bm k}/2)$, $c'=\cos(\theta_{\bm k'}/2)$, $s=\sin(\theta_{\bm k}/2)$, $s'=\sin(\theta_{\bm k'}/2)$.

Furthermore, we can write
\begin{widetext}
\begin{equation}
\label{eq_M}
M_{\bm k' \bm k}^{\sigma^\pm} = {U_{\bm k' \bm k}(V^{\sigma^\pm}_{\bm k}-V^{\sigma^\pm}_{\bm k'}) \over \hbar \omega} =
{iev_0\abs{\bm E}U_0\over \omega\sqrt{2}} \qty(cc'{\rm e}^{i\phi/2} + ss'{\rm e}^{-i\phi/2}) \qty(\sin{\theta_{\bm k}}{\rm e}^{\pm i\varphi_{\bm k}}-\sin{\theta_{\bm k'}}{\rm e}^{\pm i\varphi_{\bm k'}}),
\end{equation}
and, using the condition $2\varepsilon_k \gg \hbar \omega$, get for $\delta M_{\bm k' \bm k}^{\sigma^\pm} = 
\qty[U^{cv}_{\bm k' \bm k}V^{vc}_{\bm k}({\sigma^\pm})+V^{cv}_{\bm k'}({\sigma^\pm})U^{vc}_{\bm k' \bm k}]/( 2\varepsilon_k)$ the following expression
\begin{equation}
\label{delta_M}
\delta M_{\bm k' \bm k}^{\sigma^\pm} =
-{iev_0\abs{\bm E}U_0\over 2\sqrt{2}\varepsilon_k\omega} 
\qty[
(\cos{\theta_{\bm k}}\mp 1){\rm e}^{\pm i\varphi_{\bm k}} \qty(sc'{\rm e}^{i\phi/2} - cs'{\rm e}^{-i\phi/2})
+
(\cos{\theta_{\bm k'}}\pm 1){\rm e}^{\pm i\varphi_{\bm k'}} \qty(cs'{\rm e}^{i\phi/2} - sc'{\rm e}^{-i\phi/2})].
\end{equation}
Multiplying~\eqref{delta_M} by complex conjugate of~\eqref{eq_M} we obtain
\begin{equation}
M_{\bm k' \bm k}^*\delta M_{\bm k' \bm k} = -n_i\abs{ev_0U_0\bm E\over \omega}^2{\Phi\over 4\varepsilon_k \hbar \omega},
\end{equation}
where
\begin{multline}
\label{QM_Phi}
\Phi = \qty(cc'{\rm e}^{-i\phi/2} + ss'{\rm e}^{i\phi/2}) \qty(\sin{\theta_{\bm k}}{\rm e}^{- i\varphi_{\bm k}}-\sin{\theta_{\bm k'}}{\rm e}^{- i\varphi_{\bm k'}})
\\
\times \qty[
(\cos{\theta_{\bm k}}\mp 1){\rm e}^{\pm i\varphi_{\bm k}} \qty(sc'{\rm e}^{i\phi/2} - cs'{\rm e}^{-i\phi/2})
+
(\cos{\theta_{\bm k'}}\pm 1){\rm e}^{\pm i\varphi_{\bm k'}} \qty(cs'{\rm e}^{i\phi/2} - sc'{\rm e}^{-i\phi/2})].
\end{multline}
Next, from Eq.~\eqref{j_QM} we derive the ``interference'' contribution to the CPGE current
\begin{multline}
\bm j^{\rm QM,2} = e\sum_{\bm k, \bm k'} {2\pi\over \hbar}2{\rm Re}\qty({M_{\bm k' \bm k}^*\delta M_{\bm k' \bm k}})\delta(\varepsilon_k + \hbar\omega-\varepsilon_{k'})\qty[f_0(\varepsilon_k) - f_0(\varepsilon_{k'})] \qty[\bm v_{\bm k'}\tau(\varepsilon_{k'})-\bm v_{\bm k}\tau(\varepsilon_{k})]
\\
= -e^3{2\pi\over \hbar}n_i\abs{v_0U_0\bm E\over \omega}^2\sum_{\bm k, \bm k'} {\Phi\over 2\varepsilon_k \hbar \omega}\delta(\varepsilon_k + \hbar\omega-\varepsilon_{k'})\qty[f_0(\varepsilon_k) - f_0(\varepsilon_k + \hbar\omega)] v_0 \qty[{\bm k' \over k'}\tau(\varepsilon_k + \hbar\omega)- {\bm k\over k}\tau(\varepsilon_{k})].
\end{multline}
Since $\varepsilon_k \gg \hbar \omega$, we have $\varepsilon_{k'}\approx \varepsilon_k$ which gives for $\bm \varkappa \parallel z$
\begin{equation}
\label{j_QM_2}
j_z^{\rm QM,2} = -(ev_0)^3{2\pi\over \hbar}n_i\abs{U_0\bm E\over \omega}^2\sum_{\bm k, \bm k'} {\Phi\over 2\varepsilon_k}\delta(\varepsilon_k -\varepsilon_{k'})\qty[-f_0'(\varepsilon_k)] \qty[\cos\theta_{\bm k'}\tau(\varepsilon_k + \hbar\omega)- \cos\theta_{\bm k}\tau(\varepsilon_{k})].
\end{equation}
Averaging over directions of $\bm k$ and $\bm k'$
for the $\sigma^+$ polarization leads to
\begin{equation}
\expval{\Phi \cos\theta_{\bm k'}} = \expval{\Phi \cos\theta_{\bm k}} = -{2\over 9}.
\end{equation}
Therefore the last square brackets is proportional to $\tau(\varepsilon_k + \hbar\omega)-\tau(\varepsilon_{k})\approx -2\tau(\varepsilon_k)\hbar\omega/\varepsilon_k$, where we used the relation $\tau(\varepsilon_k) \propto \varepsilon_k^{-2}$, see Eq.~\eqref{Uquad_tau}.
Then Eq.~\eqref{j_QM_2} is reduced to
\begin{equation}
j_z^{\rm QM,2} = -{2\hbar\over 9\omega} (ev_0)^3\abs{\bm E}^2\sum_{\bm k} {\qty[-f_0'(\varepsilon_k)]\over \varepsilon_k^2}\tau(\varepsilon_k)\sum_{\bm k'}{2\pi\over \hbar}n_i\abs{U_0}^2\delta(\varepsilon_k -\varepsilon_{k'}) .
\end{equation}
The sum over ${\bm k}'$ equals to $3/\tau(\varepsilon_k)$, and we finally obtain
\begin{equation}
j_z^{\rm QM,2} =-{2\hbar\over 3\omega} (ev_0)^3\abs{\bm E}^2\sum_{\bm k} {\qty[-f_0'(\varepsilon_k)]\over \varepsilon_k^2}
=-{4\over 3}{e^3 \over h^2\omega} \abs{\bm E}^2.
\end{equation}
\end{widetext}

\section{Absence of the interband-coherence effect induced by the electric field in the course of scattering}
\label{App:ib-coherence}

In the first order in the perturbation $\delta \mathcal H=-e\bm E(t) \cdot \bm r$, where ${\bm E}(t)$ is real,  the eigenfunctions of the Hamiltonian $\mathcal H +\delta \mathcal H$ are given by
\begin{equation}
u_{c\bm{k}} = \hat{C}_{c, \bm{k}} - \frac{e \bm{E} \cdot \bm{r}_{vc}(\bm{k})}{2 \hbar v_0 k} \hat{C}_{v, \bm{k}}
= \hat{C}_{c, \bm{k}} - i \frac{e \bm{E} \cdot \bm{v}_{vc}(\bm{k})}{4 \hbar (v_0 k)^2} \hat{C}_{v, \bm{k}}.
\end{equation}
Here $\hat{C}_{c, \bm{k}} \equiv u_{\bm k}$ and $\hat{C}_{v, \bm{k}} \equiv u_{\bm k}^v$ are the unperturbed envelopes~\eqref{envelopes}, and the velocity operator $\bm v = v_0\bm \sigma$.
Then we obtain
\begin{align}
u^\dagger_{c\bm{k}'} u_{c\bm{k}} = \hat{C}^\dagger_{c, \bm{k}'} \hat{C}_{c, \bm{k}} &- \hat{C}^\dagger_{c, \bm{k}'} \hat{C}_{v, \bm{k}} i \frac{e \bm{E} \cdot \bm{v}_{vc}(\bm{k})}{4 \hbar (v_0 k)^2} 
\\
&+ \hat{C}^\dagger_{v, \bm{k}'} \hat{C}_{c, \bm{k}} i \frac{e \bm{E} \cdot \bm{v}^{*}_{vc}(\bm{k}')}{4 \hbar (v_0 k)^2},\nonumber
\end{align}
and, hence,
\begin{align}
\delta^{\bm E}\abs{\braket{u_{\bm k'}}{u_{\bm k}}}^2
= -\frac{e}{2\hbar v_0 k^2} 
\qty(R+R'),
\end{align}
where
$R$ and $R'$ are presented as
\begin{align}
R&= - \Im \Tr \left[ (\hat{C}_{c,\bm{k}'} \hat{C}^\dagger_{c,\bm{k}'}) (\hat{C}_{v,\bm{k}} \hat{C}^\dagger_{v,\bm{k}})(\bm{E} \cdot \bm{\sigma})(\hat{C}_{c,\bm{k}} \hat{C}^\dagger_{c,\bm{k}}) \right], \nonumber \\
R' &=  \Im \Tr \left[ (\hat{C}_{c,\bm{k}'} \hat{C}^\dagger_{c,\bm{k}'})(\bm{E} \cdot \bm{\sigma})(\hat{C}_{v,\bm{k}'} \hat{C}^\dagger_{v,\bm{k}'})(\hat{C}_{c,\bm{k}} \hat{C}^\dagger_{c,\bm{k}}) \right], \nonumber 
\end{align}
and for brevity we replaced here ${\bm E}(t)$ by ${\bm E}$.

Using the relations
\begin{align}
\hat{C}_{c,\bm{k}} \hat{C}^\dagger_{c,\bm{k}} &= \frac{H(\bm{k}) - E_{v\bm{k}}}{E_{c\bm{k}} - E_{v\bm{k}}} = \frac{1}{2}(1 + \bm{\sigma} \cdot \bm n_{\bm{k}}), \\
\hat{C}_{v,\bm{k}} \hat{C}^\dagger_{v,\bm{k}} &= \frac{E_{c\bm{k}} - H(\bm{k})}{E_{c\bm{k}} - E_{v\bm{k}}} = \frac{1}{2}(1 - \bm{\sigma} \cdot \bm n_{\bm{k}}),
\end{align}
where $n_{\bm{k}}=\bm{k}/k$, we get
\begin{align}
&R= - {1\over 8} \nonumber\\
&\times \Im \Tr \left[ (1 + \bm{\sigma} \cdot \bm n_{\bm{k'}}) (1 - \bm{\sigma} \cdot \bm n_{\bm{k}}) (\bm{E} \cdot \bm{\sigma}) (1 - \bm{\sigma} \cdot \bm n_{\bm{k}}) \right],  \nonumber \\
&R' ={1\over 8} \nonumber\\
& \times \Im \Tr \left[ (1 + \bm{\sigma} \cdot \bm n_{\bm{k'}}) (\bm{E} \cdot \bm{\sigma}) (1 - \bm{\sigma} \cdot \bm n_{\bm{k'}}) (1 - \bm{\sigma} \cdot \bm n_{\bm{k}}) \right]. \nonumber
\end{align}
Calculation gives
\begin{equation}
R=-R'= {1\over 2}[\bm n_{\bm{k}'} \times \bm n_{\bm{k}}]\cdot \bm E,
\end{equation}
and $\delta^{\bm E}\abs{\braket{u_{\bm k'}}{u_{\bm k}}}^2=0$.

\begin{widetext}
\section{Intraband CPGE in Weyl semimetals at anisotropic scattering}
\label{App:anis_scatt}

The equations for CPGE are obtained from those derived for isotropic scattering by a substitution 
$$n_i\abs{U_0}^2 \to \mathcal K(\abs{\bm k'-\bm k}),$$ 
where $\mathcal K(q)$ is the Fourier-component of the disorder potential correlator.
Below we consider anisotropic scattering with a Gaussian correlator:
\begin{equation}
\label{Gauss_corr}
\mathcal K(q)=\mathcal K_0 \exp(-q^2d^2).
\end{equation}

\subsection{Semiclassical derivation}

The BCD contribution is independent of scattering details and $\gamma_{\rm BCD}=-2/3$ as before.

The side-jump contribution is calculated as follows.
The second-order correction to the distribution function is still found from Eq.~\eqref{eq_f2},
where now
\begin{equation}
{1\over \tau} = \sum_{\bm k'}W_{\bm k' \bm k}^{(0)}\qty(1-\bm n_{\bm k} \cdot \bm n_{\bm k'}), \quad 
W^{(0)}_{\bm k' \bm k}={2\pi \over \hbar}\mathcal K_{\bm k' \bm k}\delta(\varepsilon_k-\varepsilon_{k'}),
\quad \mathcal K_{\bm k' \bm k}=\mathcal K(\abs{\bm k'-\bm k})\abs{\braket{u_{\bm k'}}{u_{\bm k}}}^2,
\end{equation}
with $\bm n_{\bm k}=\bm k/k$, and
\begin{equation}
\label{W_sj_anisotr}
W^{\rm sj}_{\bm k' \bm k}={2\pi \over \hbar}\mathcal K_{\bm k' \bm k} e\bm E(t)\cdot \bm r_{\bm k' \bm k} \partial_{\varepsilon_k}\delta(\varepsilon_k-\varepsilon_{k'}), \quad
\mathcal K_{\bm k' \bm k}\bm r_{\bm k' \bm k} = \mathcal K(\abs{\bm k'-\bm k}) [\bm n_{\bm k'} \times \bm n_{\bm k}] {k'+k\over 4k'k}.
\end{equation}
Since $\abs{\braket{u_{\bm k'}}{u_{\bm k}}}^2=(1+\bm n_{\bm k} \cdot \bm n_{\bm k'})/2$, we have for the energy-dependent transport scattering rate
\begin{equation}
{1\over \tau(\varepsilon_k)}={\pi \over \hbar} \nu(\varepsilon_k) \expval{\mathcal K(q_\Theta)\sin^2{\Theta}}, \quad q_\Theta =\sqrt{2}k\sqrt{1-\cos\Theta},
\end{equation}
where $\Theta$ is an angle between $\bm k$ and $\bm k'$.

The `outgoing' term in Eq.~\eqref{eq_f2} determined by $f_{\bm k}$ vanishes after averaging over all directions of $\bm k'$ 
because the argument $\abs{\bm k'-\bm k}$ depends on $\cos\Theta$ and $[\bm n_{\bm k'} \times \bm n_{\bm k}] \propto \sin{\Theta}$, see Eq.~\eqref{W_sj_anisotr}. Therefore,
the solution of Eq.~\eqref{eq_f2} can be cast in the form
\begin{equation}
f_{\bm k}^{(2)}=\tau(\varepsilon_k)\sum_{\bm k'} W^{\rm sj}_{\bm k' \bm k} f_{\bm k'}^{(1)}
=\tau(\varepsilon_k)\abs{\bm E}^2 
{e^2  v_0 \over \omega } {2\pi \over \hbar}\sum_{\bm k'}  [- f_0'(\varepsilon_{k'})] \delta'(\varepsilon_k-\varepsilon_{k'})\mathcal K(\abs{\bm k'-\bm k}) \qty(\bm \varkappa \cdot [\bm n_{\bm k'} \times[\bm n_{\bm k'} \times \bm n_{\bm k}]]) {k'+k\over 4k'k}.
\end{equation}

Then we obtain for the side-jump CPGE current density:
\begin{multline}
\bm j^{\rm sj} = ev_0\sum_{\bm k} \bm n_{\bm k} f_{\bm k}^{(2)}
={e^3  v_0^2 \over \omega }\abs{\bm E}^2\sum_{\bm k,\bm k'} 
 {2\pi \tau(\varepsilon_k) \over \hbar} [- f_0'(\varepsilon_{k'})]{k'+k\over 4k'k}\delta'(\varepsilon_k-\varepsilon_{k'}) \mathcal K(\abs{\bm k'-\bm k}) \bm n_{\bm k} \qty(\bm \varkappa \cdot [\bm n_{\bm k'} \times[\bm n_{\bm k'} \times \bm n_{\bm k}]])
\\ 
= \bm \varkappa \abs{\bm E}^2{e^3  v_0^2 \over \omega }\sum_{\bm k,\bm k'} 
 {2\pi \tau(\varepsilon_k) \over \hbar} [- f_0'(\varepsilon_{k'})]{k'+k\over 4k'k} \delta'(\varepsilon_k-\varepsilon_{k'})\mathcal K(\abs{\bm k'-\bm k}) [n_{\bm k,z} n_{\bm k',z}(\bm n_{\bm k'} \cdot \bm n_{\bm k})-n_{\bm k,z}^2 ],
\end{multline}
where we oriented the $z$ axis along $\bm \varkappa$.
Owing to the spherical symmetry of the Weyl model we can replace the variable $z$ with $x$ and $y$ in this expression, and then sum up three such terms dividing the result by three. This transforms $n_{\bm k,z} n_{\bm k',z} \to (\bm n_{\bm k'} \cdot \bm n_{\bm k})$ and $n_{\bm k,z}^2 \to 1$.
As a result we get
%
\begin{equation}
\bm j^{\rm sj} =-{2e^3  v_0^2 \over 3\omega }\bm \varkappa \abs{\bm E}^2\int \dd \varepsilon_k \int \dd \varepsilon_{k'} \nu(\varepsilon_{k'})
 [- f_0'(\varepsilon_{k'})]{k'+k\over 4k'k} \delta'(\varepsilon_k-\varepsilon_{k'}){\expval{\mathcal K(q_{\bm k'\bm k}) \sin^2\Theta}\over \expval{\mathcal K(q_\Theta)\sin^2{\Theta}}},
\end{equation}
where $q_{\bm k'\bm k}=\sqrt{k'^2+k^2-2kk'\cos\Theta}$.
By using $1/k=\hbar v_0/\varepsilon_k$ and $\delta'(\varepsilon_k-\varepsilon_{k'})=-\partial_{\varepsilon_{k'}}\delta(\varepsilon_k-\varepsilon_{k'})$, we integrate over $\varepsilon_{k'}$ by parts and obtain
\begin{multline}
\bm j^{\rm sj} =-{e^3  \hbar v_0^3 \over 6\omega }\bm \varkappa \abs{\bm E}^2\int \dd \varepsilon_{k'} \lim_{\varepsilon_k \to  \varepsilon_{k'}}\partial_{\varepsilon_{k'}}
\qty[- f_0'(\varepsilon_{k'})\nu(\varepsilon_{k'})\qty({1\over \varepsilon_k}+{1\over \varepsilon_{k'}}) {\expval{\mathcal K(q_{\bm k'\bm k}) \sin^2\Theta}\over \expval{\mathcal K(q_\Theta)\sin^2{\Theta}}}]
\\
=-{e^3  \hbar v_0^3 \over 6\omega }\bm \varkappa \abs{\bm E}^2\int \dd \varepsilon_{k'} \lim_{\varepsilon_k \to  \varepsilon_{k'}}
\qty{
\partial_{\varepsilon_{k'}} \qty[- f_0'(\varepsilon_{k'})\nu(\varepsilon_{k'})\qty({1\over \varepsilon_k}+{1\over \varepsilon_{k'}}) ]
+\qty[- f_0'(\varepsilon_{k'})\nu(\varepsilon_{k'}){2\over \varepsilon_{k'}}]\partial_{\varepsilon_{k'}} {\expval{\mathcal K(q_{\bm k'\bm k}) \sin^2\Theta}\over \expval{\mathcal K(q_\Theta)\sin^2{\Theta}}}}.
\end{multline}
Here the first term is calculated in the same way as at isotropic scattering:
\begin{multline}
\int \dd \varepsilon_{k'} \lim_{\varepsilon_k \to  \varepsilon_{k'}}\partial_{\varepsilon_{k'}} \qty[- f_0'(\varepsilon_{k'})\nu(\varepsilon_{k'})\qty({1\over \varepsilon_k}+{1\over \varepsilon_{k'}}) ]
=\int \dd \varepsilon_{k'} \qty{ {2\over \varepsilon_{k'}}\partial_{\varepsilon_{k'}} \qty[- f_0'(\varepsilon_{k'})\nu(\varepsilon_{k'}) ] + {f_0'(\varepsilon_{k'})\nu(\varepsilon_{k'}) \over \varepsilon_{k'}^2}}
\\=\int \dd \varepsilon_{k'} \qty[- f_0'(\varepsilon_{k'}) ]  {\nu(\varepsilon_{k'})\over \varepsilon_{k'}^2}
={1\over 2\pi^2(\hbar v_0)^3}.
\end{multline}
Therefore we obtain 
for the coefficient $\gamma_{\rm sj}$
\begin{equation}
\gamma_{\rm sj}=-{1\over 3} \qty[1+2\varepsilon_{\rm F} {\expval{\partial_{\varepsilon_{k'}}\mathcal K(q_{\bm k'\bm k}) \sin^2\Theta}\over \expval{\mathcal K(q_\Theta)\sin^2{\Theta}}}],
\end{equation}
where averaging is performed over the Fermi sphere $k=k'=k_{\rm F}$.

For the Gaussian correlator~\eqref{Gauss_corr} we have
\begin{equation}
\varepsilon_{\rm F}\partial_{\varepsilon_{k'}}\mathcal K(q_{\bm k'\bm k})|_{k=k'=k_{\rm F}}
=-{2k_{\rm F}^2d^2\over \hbar v_0}\mathcal K(q_\Theta)(1-\cos\Theta) =-\alpha \mathcal K_0 {\rm e}^{-\alpha} \exp(\alpha \cos{\Theta})(1-\cos\Theta),
\end{equation}
where $\alpha = 2(k_{\rm F}d)^2$.
This gives
\begin{multline}
\gamma_{\rm sj}=-{1\over 3} \qty[1-2\alpha  {\expval{\exp(\alpha \cos{\Theta})(1-\cos\Theta) \sin^2\Theta}\over \expval{\exp(\alpha \cos{\Theta})\sin^2{\Theta}}}]
=-{1\over 3} \qty[1-2\alpha  {\int_{-1}^1 \dd \mu \exp(\alpha \mu)(1-\mu) (1-\mu^2)\over \int_{-1}^1 \dd \mu \exp(\alpha \mu)(1-\mu^2)}]
\\
=-{1\over 3} \qty{1-2 \frac{3 (e^{2 \alpha }-1)-2 \alpha  [\alpha  (2 \alpha +3)+3]}{e^{2 \alpha } (\alpha -1)+\alpha +1}}.
\end{multline}
Finally, the semiclassical result takes the form
\begin{equation}
\label{gamma_SC_anisotr}
\gamma_{\rm SC} = \gamma_{\rm BCD}+\gamma_{\rm sj}=-1+2\frac{e^{2 \alpha }-1-2 \alpha  [\alpha  (2 \alpha/3 +1)+1]}{e^{2 \alpha } (\alpha -1)+\alpha +1}.
\end{equation}
In the limit $d \to 0, \alpha \to 0$, Eq.~\eqref{gamma_SC_anisotr} passes to Eq.~\eqref{gamma_SC_isotr}.

\subsection{Quantum-mechanical derivation}

For the intraband optical matrix elements one has
\begin{equation}
M_{\bm k' \bm k}^*\delta M_{\bm k' \bm k} = -\mathcal K(\abs{\bm k'-\bm k})\abs{ev_0\bm E\over \omega}^2{\Phi\over 4\varepsilon_k \hbar \omega},
\quad
\abs{\delta M_{\bm k' \bm k}}^2=\mathcal K(\abs{\bm k'-\bm k})\abs{ev_0\bm E\over \omega}^2{\Psi\over 8\varepsilon_k^2},
\end{equation}
where $\Phi$ is given by Eq.~\eqref{QM_Phi} and
\begin{equation}
\Psi=\abs{
(\cos{\theta_{\bm k}}\mp 1){\rm e}^{\pm i\varphi_{\bm k}} \qty(sc'{\rm e}^{i\phi/2} - cs'{\rm e}^{-i\phi/2})
+
(\cos{\theta_{\bm k'}}\pm 1){\rm e}^{\pm i\varphi_{\bm k'}} \qty(cs'{\rm e}^{i\phi/2} - sc'{\rm e}^{-i\phi/2})}^2.
\end{equation}

The CPGE current is
\begin{equation}
\label{CPGE}
j_{\rm CPGE} = \gamma_{\rm QM}{e^3 \over h^2\omega} \abs{\bm E}^2,
\quad
\gamma_{\rm QM} =  {\expval{\mathcal K(q_\Theta)\qty[{1\over 4}\Psi(\cos{\theta_{\bm k'}}-\cos{\theta_{\bm k}})+\Phi(\cos{\theta_{\bm k'}}+\cos{\theta_{\bm k}})]}
\over \expval{\mathcal K(q_\Theta){1+\cos{\Theta}\over 2}(1-\cos{\Theta})}}.
\end{equation}
The scattering angle can be expressed via $\theta_{{\bm k}'}$ and $\theta_{\bm k}$ by
\begin{equation}
\cos{\Theta}=\cos{\theta_{\bm k}}\cos{\theta_{\bm k'}} + \sin{\theta_{\bm k}}\sin{\theta_{\bm k'}} \cos\phi.
\end{equation}

At isotropic scattering, when $\mathcal K=\mathcal K_0$ is independent of $\Theta$, 
averaging yields
\begin{equation}
\expval{{1\over 4}\Psi(\cos{\theta_{\bm k'}}-\cos{\theta_{\bm k}})}={4\over 9}, \quad \expval{\Phi(\cos{\theta_{\bm k'}}+\cos{\theta_{\bm k}})}=-{4\over 9},
\quad j_{\rm CPGE}^{\rm isotr} =0.
\end{equation}

For the scattering with a Gaussian correlator~\eqref{Gauss_corr}, where ${\cal K}(q_{\Theta}) = {\cal K}_0 {\rm e}^{- \alpha} \exp (\alpha \cos{\Theta})$, we obtain for the CPGE coefficient
\begin{equation}
\gamma_{\rm QM} = {\alpha^3\over \alpha\cosh\alpha - \sinh\alpha}
\expval{\exp(\alpha \cos{\Theta})\qty[{1\over 4}\Psi(\cos{\theta_{\bm k'}}-\cos{\theta_{\bm k}})+\Phi(\cos{\theta_{\bm k'}}+\cos{\theta_{\bm k}})]}.
\end{equation}
Here averaging means 
\begin{equation}
\expval{\mathcal F(\theta_{\bm k},\theta_{\bm k'},\phi)} = {1\over8\pi}\int_0^\pi \dd \theta_{\bm k} \sin{\theta_{\bm k}}\int_0^\pi \dd \theta_{\bm k'} \sin{\theta_{\bm k'}} \int_0^{2\pi} \dd \phi \mathcal F(\theta_{\bm k},\theta_{\bm k'},\phi).\nonumber
\end{equation}

The dependencies $\gamma_{\rm SC}(\alpha)$ and numerically evaluated $\gamma_{\rm QM}(\alpha)$ are plotted in Fig.~\ref{Fig_Gauss_scatt}.

\end{widetext}

\bibliography{CPGE_WSM.bib}
\end{document}